\documentclass{article}
\usepackage{frascatiphys,here,graphicx,subfigure}
\usepackage{color}
\begin{document}
\title{PHASE STRUCTURE, CRITICAL POINTS, \\
AND SUSCEPTIBILITIES IN NAMBU-JONA-LASINIO TYPE MODELS }
\author{C. A de Sousa, Pedro Costa and M. C. Ruivo\\
{\em Departamento de F\'{\i}sica, Universidade de Coimbra, P - 3004 - 516} \\
{\em Coimbra, Portugal} }

\maketitle
\baselineskip=11.6pt
\begin{abstract}
We investigate the chiral phase transition at finite temperature and chemical potential
within SU(2) and  SU(3) Nambu-Jona-Lasinio  type  models. The behavior of the baryon
number susceptibility and the specific heat, in the vicinity of the critical end point,
is studied.  The class of the critical points is analyzed by calculating critical
exponents.
\end{abstract}
\baselineskip=14pt
\section{Introduction}

Strongly interacting matter at non-zero temperature and chemical potential is an exciting
topic for physicists coming from different areas, either theoretical or  experimental.
One of the main goals in the heavy-ion physics program nowadays is to study the effects
of several macroscopic  phenomena occurring under extreme conditions. The discussion
about the existence of a tricritical point (TCP) or a critical end point (CEP) is also a
topic of recent interest. As is well known,  a TCP separates the first order transition
at high chemical potentials from the second order transition at high temperatures. If the
second order transition is replaced by a smooth crossover, a CEP which separates the two
lines is found.
At the CEP the phase transition is of second order  and probably falls into  the same
universality class of the three-dimensional Ising model.
The existence of the CEP in QCD was suggested at the end of the
eighties\cite{Asakawa:1989NPA}, and its properties in the context of several models have
been studied since then\cite{Hatta:2003PRD,Schaefer:2006,Costa:2007PLB}.
The most recent lattice results with $N_f=2+1$ staggered quarks of physical masses
indicate the location of the CEP at $T^{CEP}=162\pm2\,\mbox{MeV}$ and
$\mu^{CEP}=360\pm40\,\mbox{MeV}$\cite{Fodor:2004JHEP}, however its exact location is not
yet known.

This point of the phase diagram is the special focus of the present contribution.
Nambu-Jona-Lasinio (NJL) type models  are  used and the main goal is to locate the
critical end point and confront the results with universality arguments.

We remark that most of the work done in this area has been performed with non strange
quarks only.
We will discuss the class of the critical points by including the analyzes in the chiral
limit  of both SU(2) and SU(3) versions of the NJL model.

\vskip0.2cm

The Lagrangian of the SU(3) NJL model\cite{Kunihiro:PR1994,Rehberg:1995PRC} is given by:
\begin{eqnarray} \label{lagr}
{\mathcal L} &=& \bar{q} \left( i \partial \cdot \gamma - \hat{m} \right) q
+ \frac{g_S}{2} \sum_{a=0}^{8} \Bigl[ \left( \bar{q} \lambda^a q \right)^2+ \left(
\bar{q} (i \gamma_5)\lambda^a q \right)^2
 \Bigr] \nonumber \\
&+& g_D \Bigl[ \mbox{det}\bigl[ \bar{q} (1+\gamma_5) q \bigr]
  +  \mbox{det}\bigl[ \bar{q} (1-\gamma_5) q \bigr]\Bigr] \, .
\end{eqnarray}

Using a standard hadronization procedure\cite{Costa:2003PRC,Costa:2005PRD}, the baryonic thermodynamic potential,
$\Omega(T, V,\mu_i)$, is  obtained directly from the effective action.
The baryon number susceptibility $\chi_B$  and the specific heat $C$ describe,
respectively,  the response of the baryon density $\rho_B$ and the entropy $S$ with
respect to the chemical potential $\mu_i$ and the temperature $T$:
\begin{equation}
    \chi_B = \frac{1}{3}\sum_{i=u,d,s}\left(\frac{\partial
    \rho_i}{\partial\mu_i}\right)_{T} \hskip0.3cm {\rm and} \hskip0.3cm
     C = \frac{T}{V}\left ( \frac{\partial S}{\partial T}
    \right)_{N_i}.
\end{equation}

These physical quantities are relevant observables to be studied  in the context of
possible signatures for chiral symmetry restoration.

Our model of strong interacting matter can simulate   regions   of  a hot and dense
fireball created in a heavy-ion collision.  Since electrons and positrons are not
involved in the strong interaction,  we impose the condition $\mu_e=0$. So, we naturally
get the chemical equilibrium condition $\mu_u=\mu_d=\mu_s=\mu_B$ that will be used.

After this presentation  of the model, we discuss the
phase diagrams in sec. 2. The behavior of the baryon number susceptibility and the
specific heat  in the $T-\mu_B$ plane around the CEP is studied in sec. 3,  as well as
the corresponding critical exponents. Finally, we conclude in sec. 4 with a brief summary
of our results.

\section{Phase diagrams in SU(2) and SU(3) NJL models}

In this section we analyze  the phase diagrams  in different conditions in the $T-\mu_B$
plane. Depending on the number of quark flavors $N_f=2$ or $N_f=3$, and on the masses of
the quarks, different situations can occur and the transition from hadronic matter to QGP
may be of first order, second order, or a crossover transition.

\begin{figure}[H]
    \begin{center}
        {\includegraphics[scale=0.20]{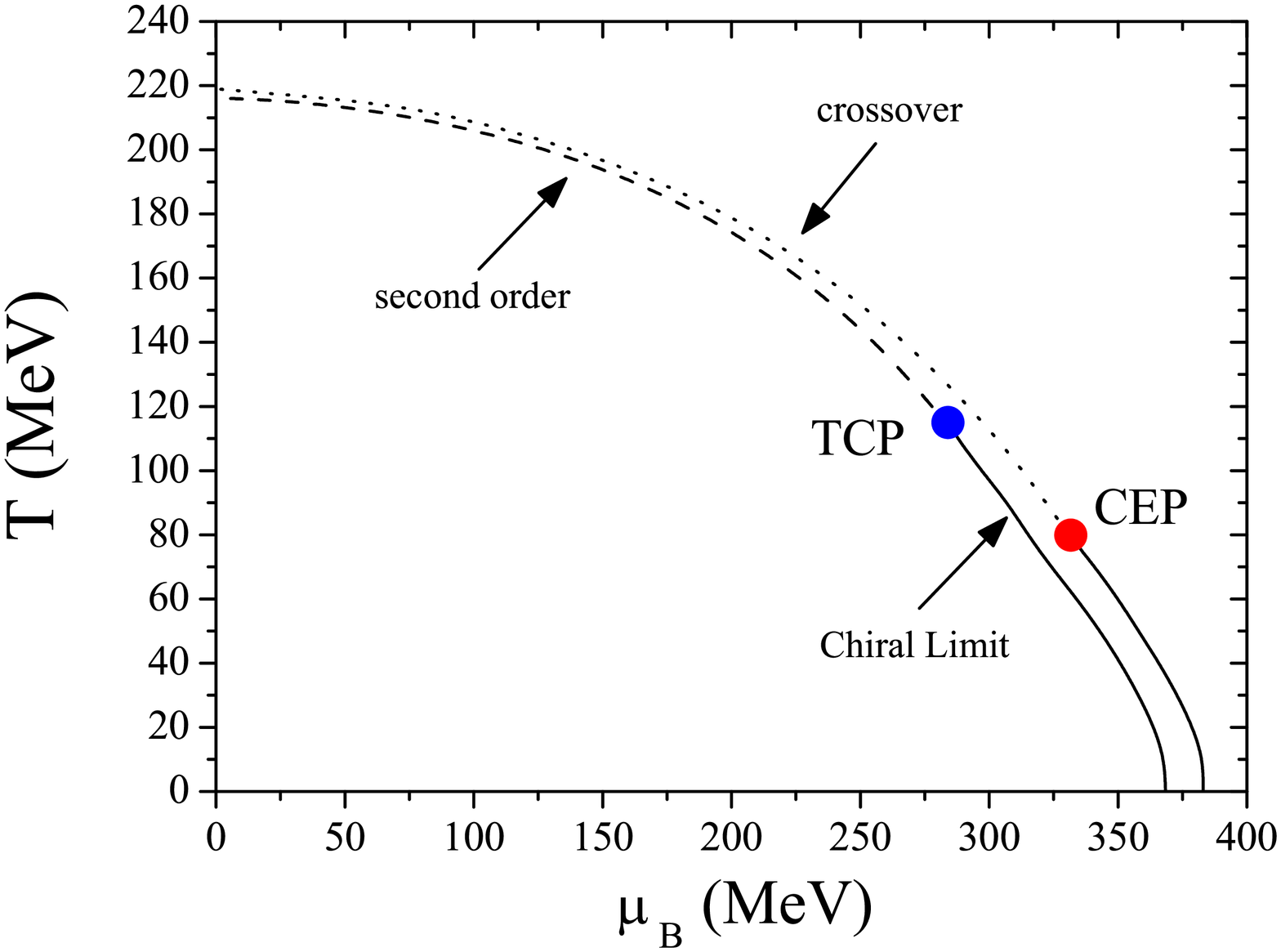}}
        {\includegraphics[scale=0.20]{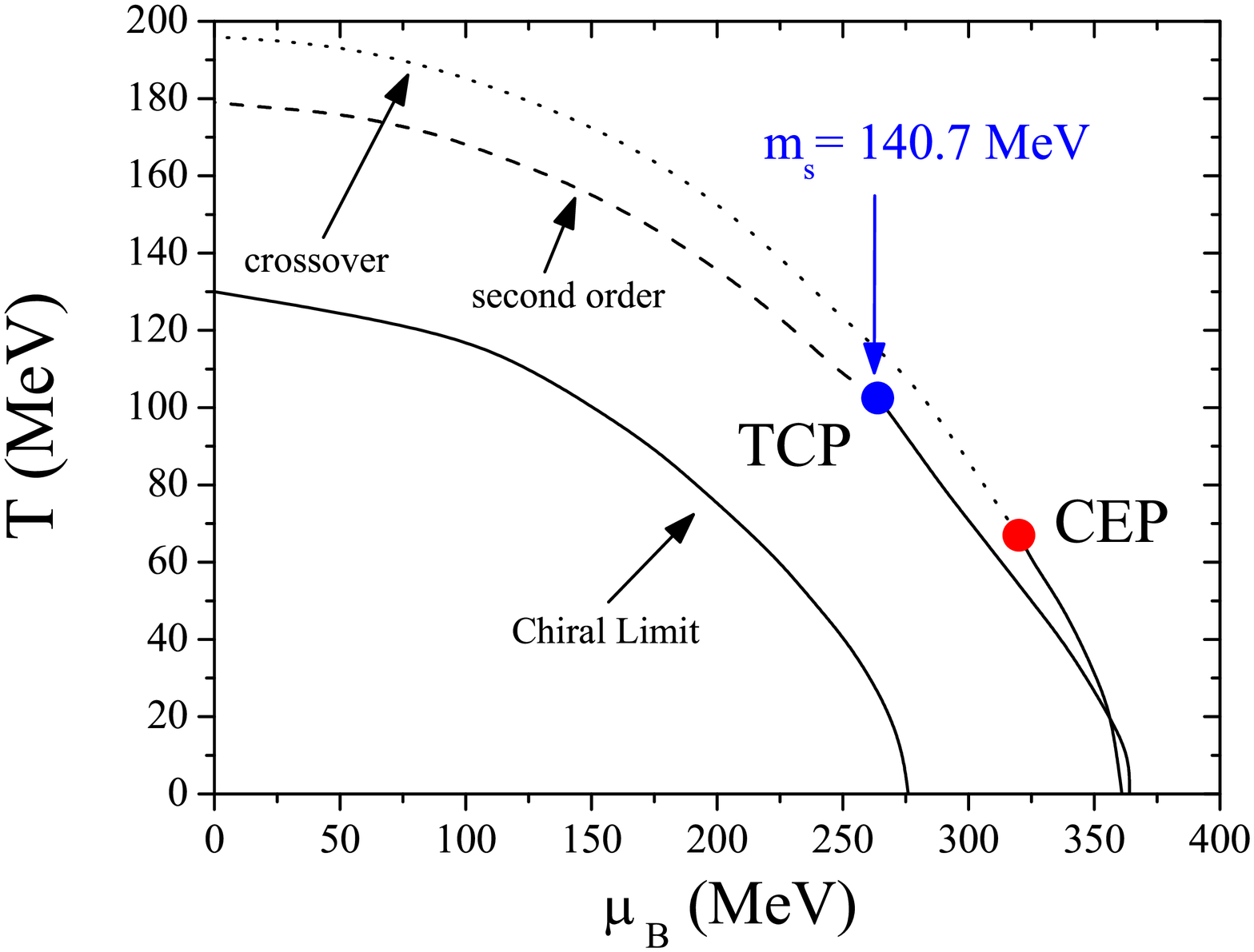}}
\vspace{-.5cm}
        \caption{\it Phase diagram in the SU(2) (left) and SU(3) (right) NJL models.
        The solid line represents the first order transition, the dashed the second 
        order and the dotted the crossover transition.} 
        \label{Fig:diagfases}
    \end{center}
\end{figure}

We start by analyzing  the differences between the three-flavor NJL model
and its simpler version in the SU(2) sector. The phase diagrams for both models  are
presented in fig. \ref{Fig:diagfases} as a function of $\mu_B$ and $T$.

Concerning the SU(2) model, and  using physical values of the quark masses, $m_u = m_d =
6$ MeV, together with $\Lambda=590$ MeV and $g_S\,\Lambda^2=2.435$, we find that the CEP
is localized at $T^{CEP}=79.9$ MeV and $\mu_B^{CEP} = 331.72$ MeV.
We  verified that, in the chiral limit,  the transition is of second order at $\mu_B=0$
and, as $\mu_B$ increases, the line of  second order phase transition will end in a first
order line at the TCP. The TCP is located at $\mu_B^{TCP}=286.1$ MeV and $T^{TCP}=112.1$
MeV.

\vspace{-.5cm}

\vspace{-.5cm} 
For the SU(3) NJL model, also in the chiral limit ($m_u=m_d=m_s=0$), we
verify that the phase diagram does not exhibit a TCP: chiral symmetry is restored via a
first order transition for all baryonic chemical potentials and temperatures (see right
panel of fig. \ref{Fig:diagfases}). This pattern of chiral symmetry restoration remains
for $m_u=m_d=0$ and $m_s<m_{s}^{crit}$. In our model we found $m_{s}^{crit}=18.3$ MeV for
$m_u=m_d=0$\cite{Costa:2007PLB}. When $m_s\geq m_{s}^{crit}$, at $\mu_B=0$, the
transition is second order and, as $\mu_B$ increases, the  second order line  will end in
a first order line at the TCP.  The TCP for $m_{s}=140.7$ MeV  is located at
$\mu_B^{TCP}=265.9$ MeV and $T^{TCP}=100.5$ MeV. If we choose $m_u=m_d\neq0$, instead of
second order transition we have a smooth crossover which critical line will end in the
first order line at the CEP.
Using the set of parameters\cite{Rehberg:1995PRC,Costa:2005PRD}: $m_u = m_d = 5.5$ MeV,
$m_s = 140.7$ MeV,  $g_S \Lambda^2 = 3.67$, $g_D \Lambda^5 = -12.36$ and $\Lambda =
602.3$ MeV, this point is localized at $T^{CEP}=67.7$ MeV and $\mu_B^{CEP} = 318.4$ MeV.

We point out that both situations are in agreement with what is expected at
$\mu_B=0$\cite{Pisarski:1984PRD}: the  phase transition in the chiral limit is of second
order for $N_f = 2$ and first order for $N_f\geq3$.

\section{Susceptibilities and critical exponents  in the vicinity of the CEP}

The phenomenological relevance of fluctuations  around the CEP/TCP of QCD has been
recognized by several authors.

\vspace{-.5cm}
\begin{figure}[H]
    \begin{center}
        {\includegraphics[scale=0.21]{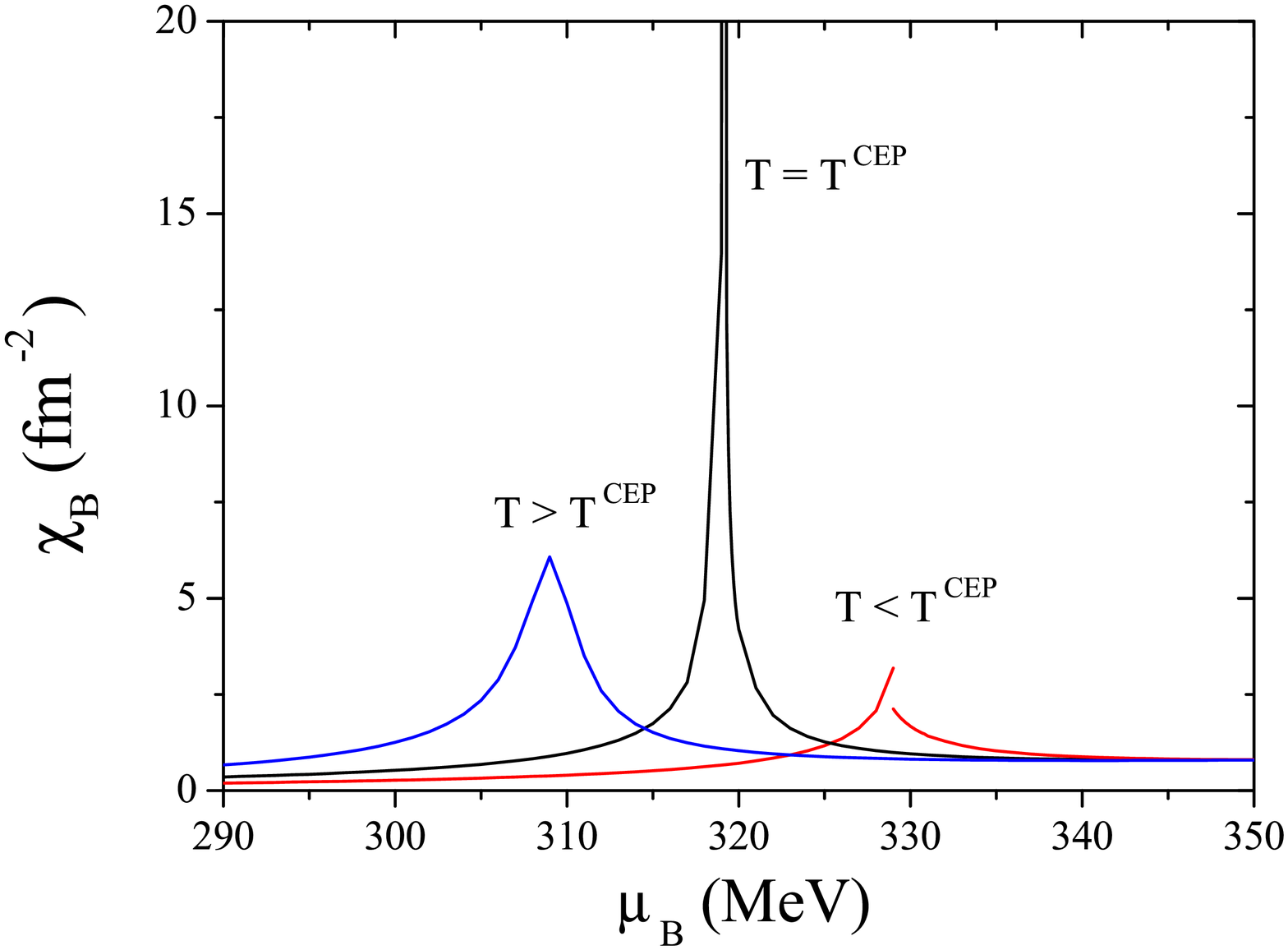}}
        {\includegraphics[scale=0.21]{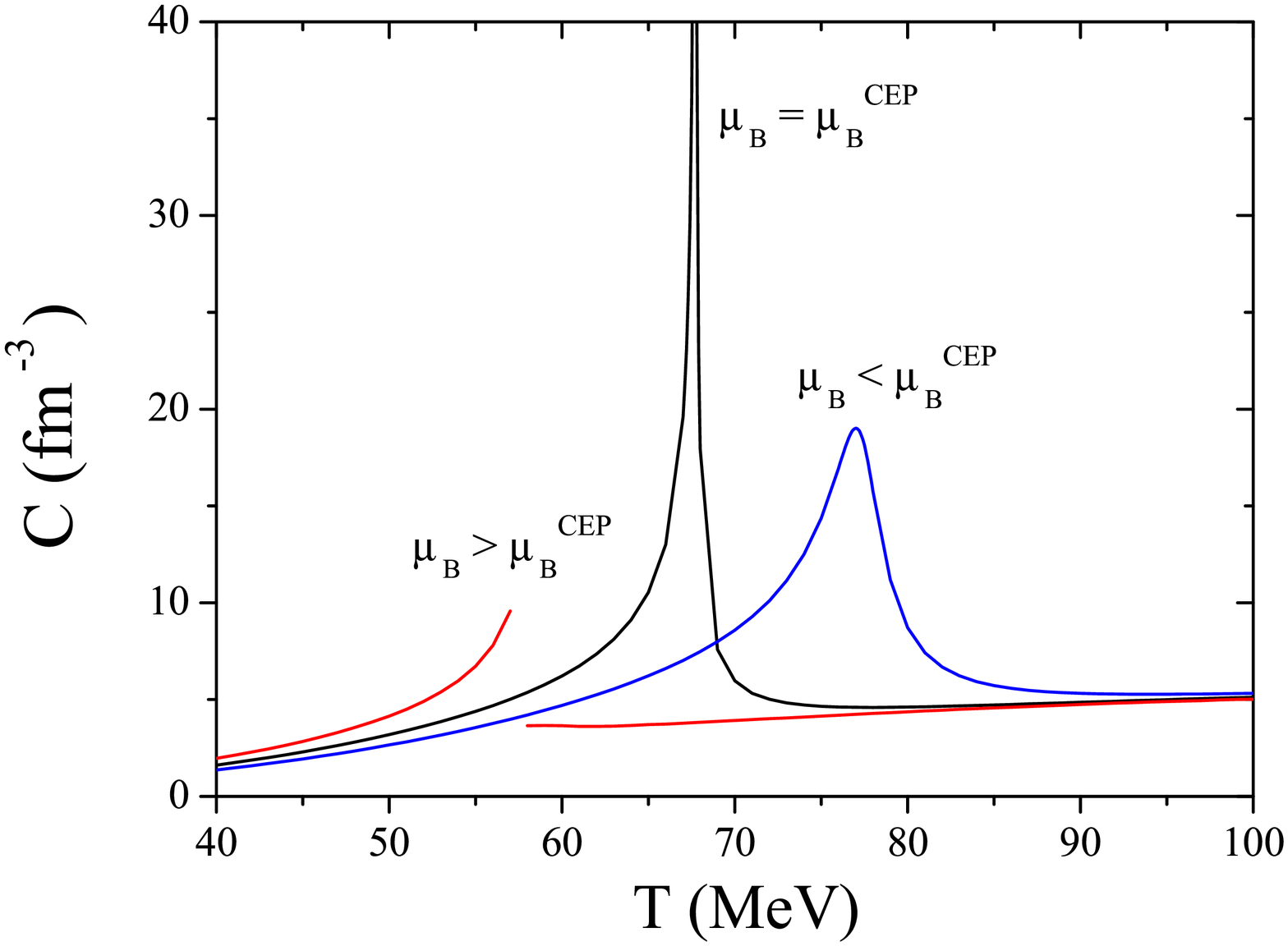}}
        \vspace{-.5cm}\caption{\it Response functions in the SU(3) NJL model. 
        Left panel: Baryon number susceptibility as a function of $\mu_B$ for 
        different $T$ around the CEP ($T^{CEP}=67.7$ MeV and$T=T^{CEP}\pm10$ MeV).
        Right panel: Specific heat as  a function of $T$ for different  
        $\mu_B$ around the CEP ($\mu_B^{CEP}=318.4$ MeV and $\mu_B=\mu_B^{CEP}\pm10$ MeV).} 
        \label{CEPchiT}
    \end{center}
\end{figure}
\vspace{-.5cm}

In the left panel of fig. \ref{CEPchiT},  the baryon number susceptibility is plotted for
three different temperatures around the CEP. For temperatures below $T^{CEP}$ the phase
transition is  first order  and, consequently, $\chi_B$ has a discontinuity. For $T =
T^{CEP}$ the susceptibility diverges at $\mu_B=\mu_B^{CEP}$ (the slope of the baryon
number density tends to infinity). For temperatures above $T^{CEP}$, in the crossover
region, the discontinuity of $\chi_B$ disappears at the transition line.
A similar behavior is found for the specific heat for three different chemical potentials
around the CEP, as we can observe from the right panel of fig. \ref{CEPchiT}. These
calculations have been performed in the SU(3) NJL model, but the same qualitative
behavior can be found in the SU(2) NJL version\cite{Costa:2007PLB}.

Summarizing, the  baryon number susceptibility and the
specific heat diverge at $T = T^{CEP}$ and $\mu\,=\,\mu^{CEP}$,
respectively.\cite{Hatta:2003PRD,Schaefer:2006,{Costa:2007PLB}} In order to make this
statement more precise, we will focus on the values of a set of indices, the so-called
critical exponents, which describe the behavior near the critical point of various
quantities of interest (in our case $\epsilon$ and $\alpha$ are the critical exponents of
$\chi_B$ and $C$, respectively).
If the critical region of the CEP is small, it is expected that most of the fluctuations
associated with the CEP will come from the mean field region around the
CEP\cite{Hatta:2003PRD}.

To a better understanding of the critical behavior of the system, we  also analyze  what
happens  in the SU(2) NJL model.

\vspace{-.5cm}

\begin{figure}[H]
    \begin{center}
        {\includegraphics[scale=0.20]{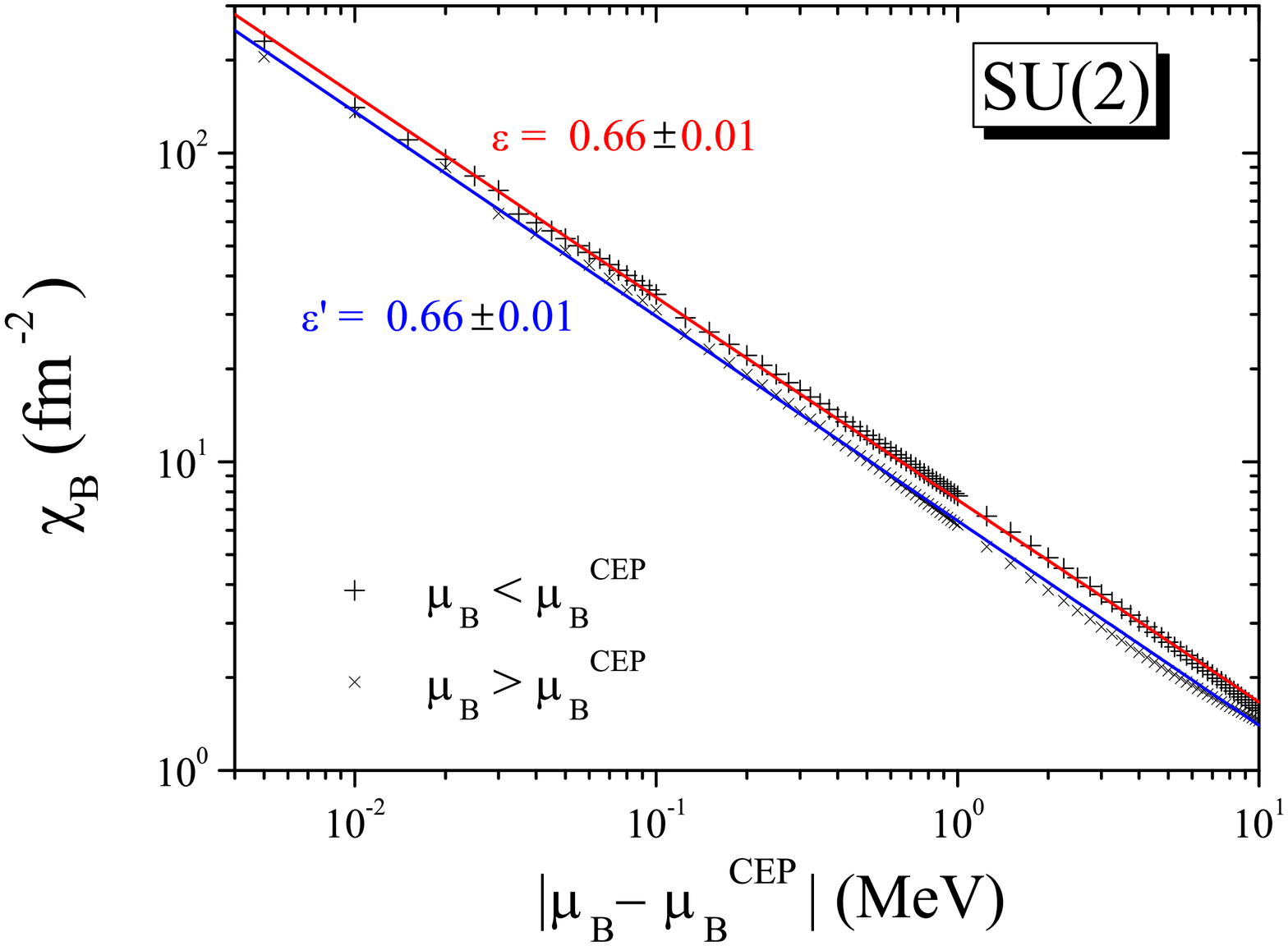}}
        {\includegraphics[scale=0.20]{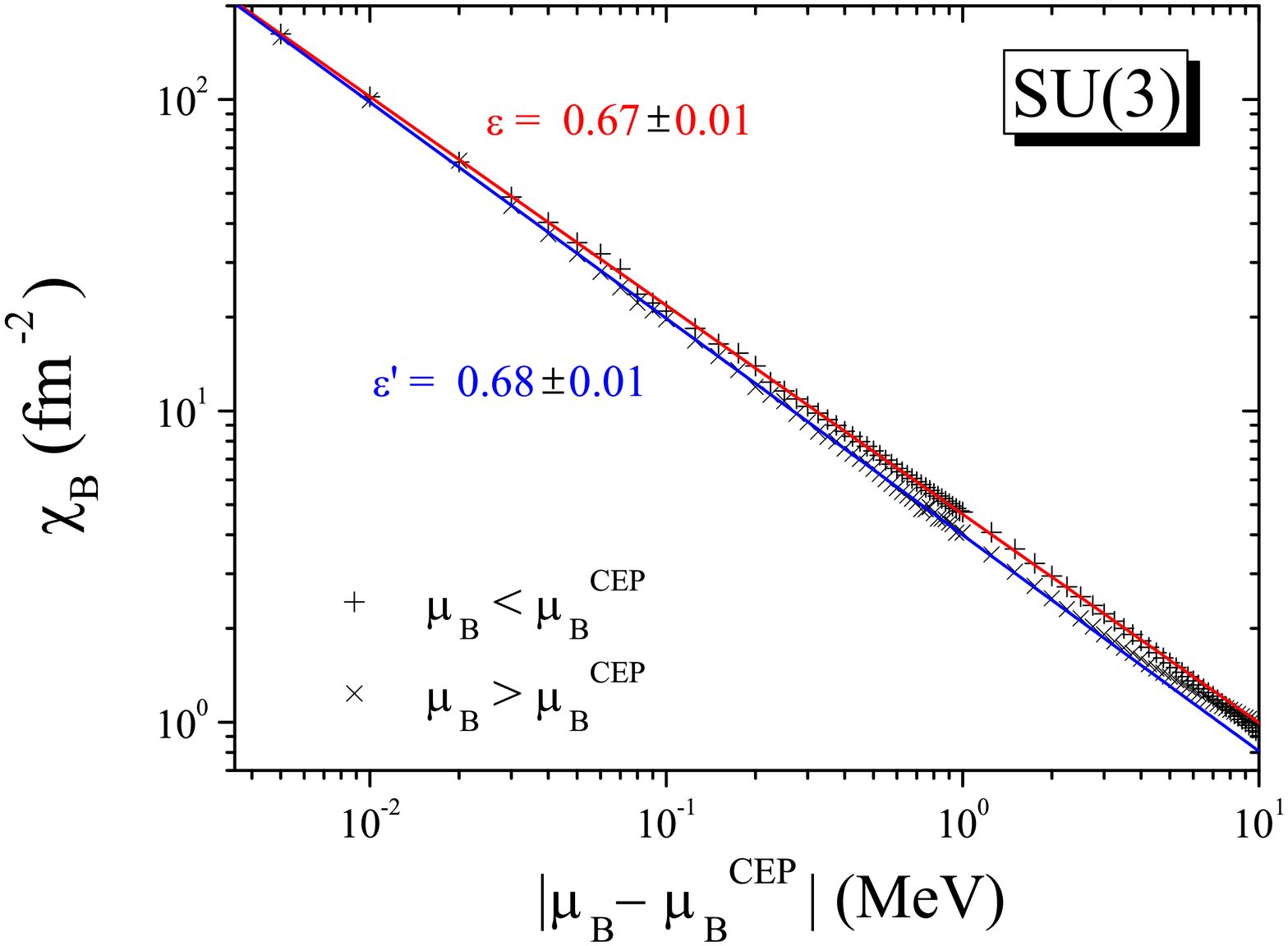}}
       \vspace{-.5cm} \caption{\it  Baryon number susceptibility as a function of 
       $|\mu_B-\mu_B^{CEP}|$ at fixed temperature $T=T^{CEP}$ in SU(2) 
       (left panel) and SU(3) (right panel) NJL models.} 
       \label{expchi23}
    \end{center}
\end{figure}

\vspace{-.5cm}

To obtain  the critical exponent $\epsilon (\epsilon^\prime$)  for the baryon number
susceptibility, we will consider   a path parallel to the $\mu_B$-axis in the $T-\mu_B$
plane, from lower (higher) $\mu_B$ towards the critical $\mu_B^{CEP}$, at fixed
temperature $T = T^{CEP}$.  To this purpose we consider  a linear logarithmic fit of the
type
$\ln \chi_B = -\epsilon^{(\prime)} \ln |\mu_B -\mu_B^{CEP} | + c^{(\prime)}_1$ ,
where the term $c_1 \,(c^\prime_1)$ is independent of $\mu_B$.

The values presented in fig.  \ref{expchi23} for these critical exponents, calculated in
both SU(2) and SU(3) NJL models, are consistent with the mean field theory prediction
$\epsilon = 2/3$.
This means that the size of the region  is approximately the same independently of the
direction of the path parallel to the $\mu_B$-axis.

\begin{figure}[H]
    \begin{center}
        {\includegraphics[scale=0.20]{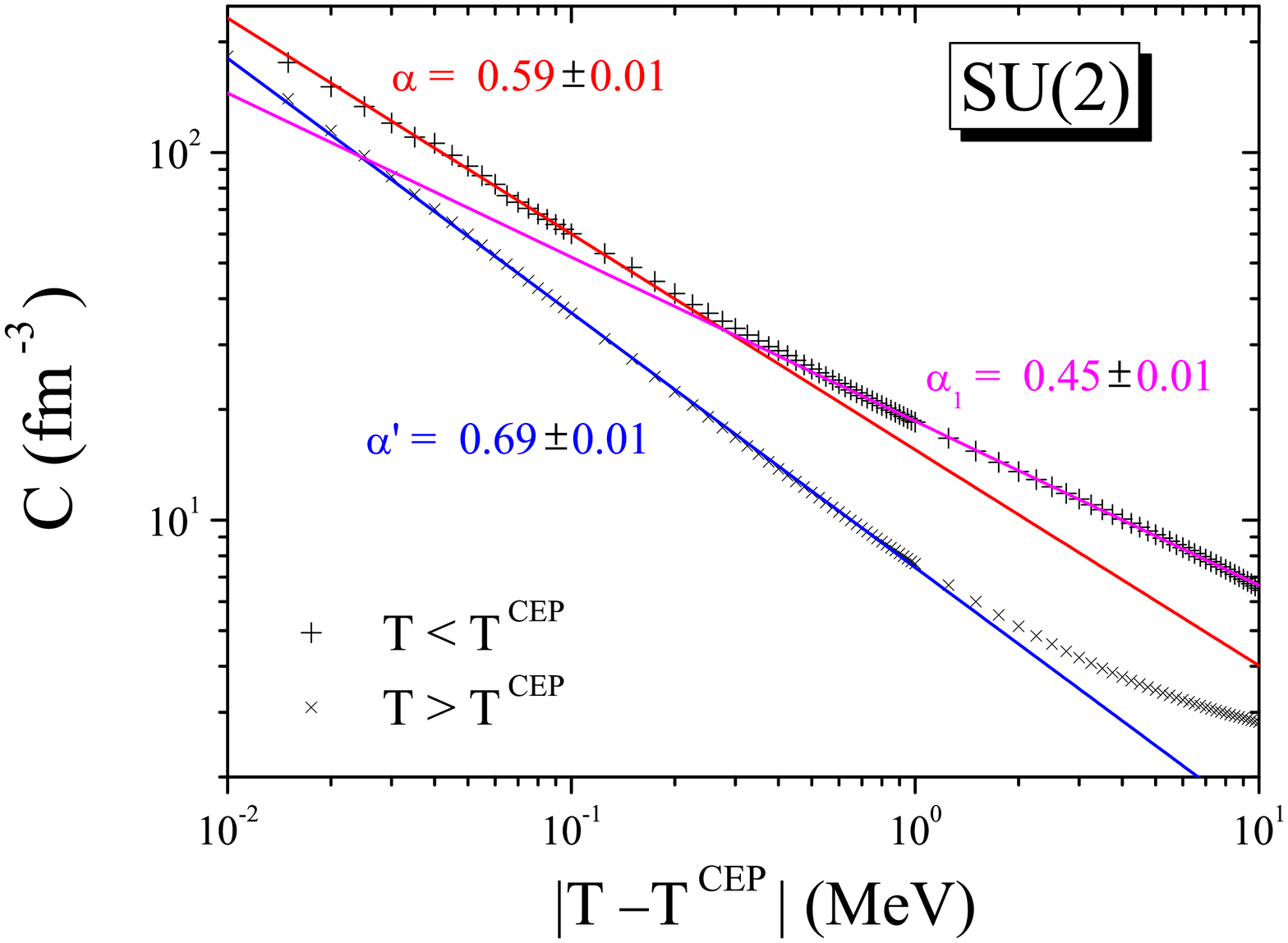}}
        {\includegraphics[scale=0.20]{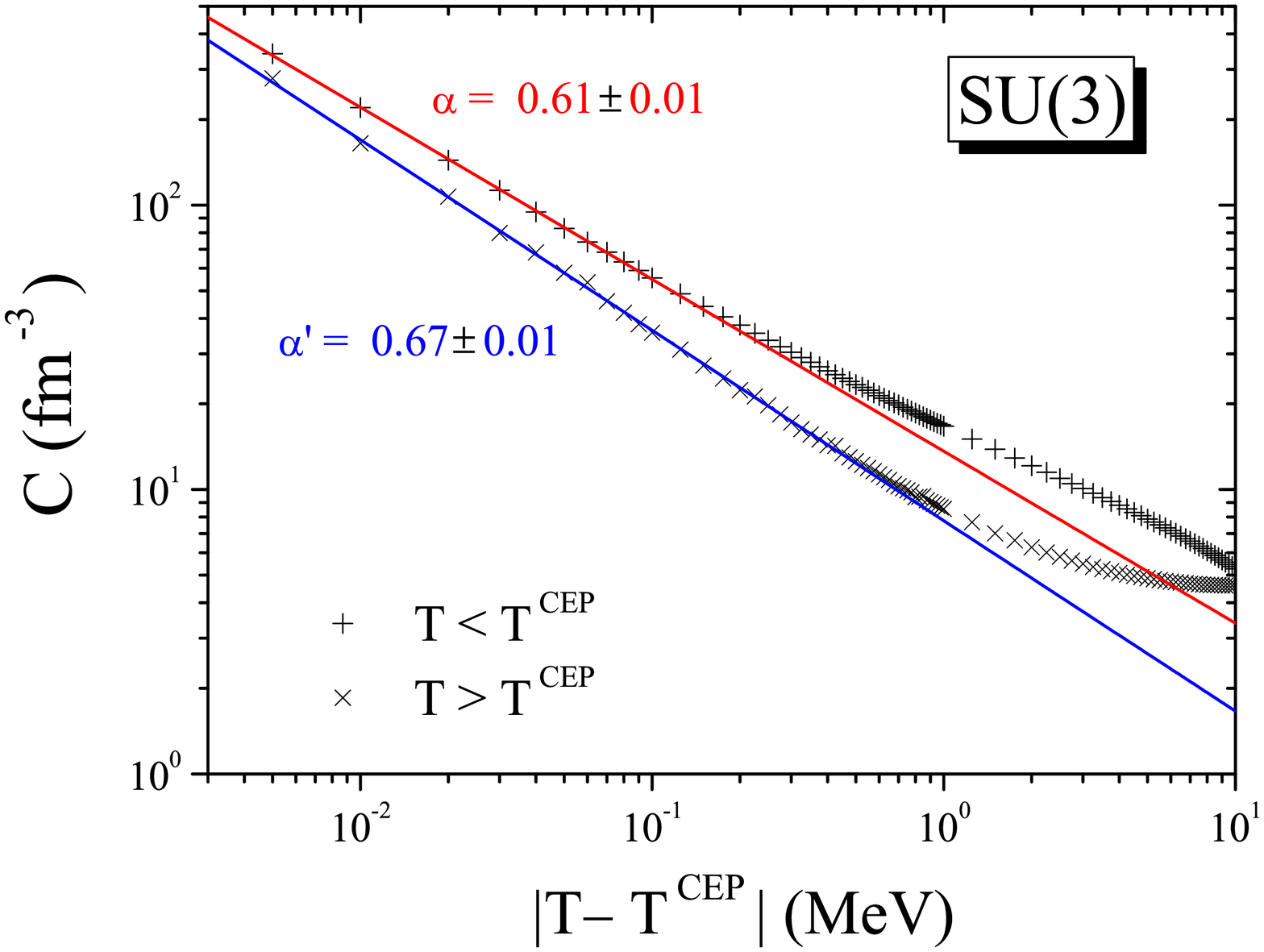}}
       \vspace{-.5cm} \caption{\it Specific heat as a function of $T$ for different values of 
       $\mu_B$ around  $\mu=\mu_B^{CEP}$ in SU(2) (left panel) and SU(3) (right panel) NJL 
       models.} 
       \label{expC23}
    \end{center}
\end{figure}

Paying now attention to the specific heat  around the CEP, we
have used a path parallel to the $T$-axis in the $T-\mu_B$ plane from lower/higher $T$
towards  $T^{CEP}$ at fixed $\mu_B = \mu_B^{CEP}$. In fig. \ref{expC23} we plot $C$ as a
function of $T$ close to the CEP in a logarithmic scale for both SU(2) and SU(3)
calculations.
In this case we use a linear logarithmic fit, $\ln C = -\alpha \ln |T -T^{CEP}| + c_2$,
where the term $c_2$ is independent of $T$.

Starting with the SU(2) case, we observe (see left panel of fig. \ref{expC23}), for
$T<T^{CEP}$, that the   slope of the fitting of data points  changes for  $|T-T^{CEP}|$
around $0.3$ MeV. So we have a change from the  critical exponent $\alpha=0.59\pm 0.01$
to $\alpha_1=0.45\pm 0.01$.
As pointed out in\cite{Hatta:2003PRD}, this change of the exponent can be interpreted as
a crossover of different universality classes, with the CEP being affected by the TCP. It
seems that the effect of the hidden TCP on the CEP is relevant  for the specific heat
contrarily to what happens to $\chi_B$.

\vspace{-.5cm}

\vspace{-.5cm}

We also observe that there is no clear evidence of  change of the slope of the fitting of
data points  in the three-flavor NJL model (see fig. \ref{expC23}, right panel).
In fact, now we  only  obtain  a critical exponent $\alpha=0.61\pm 0.01$ when the
critical point is approached from below. When the critical point is approached from above
the trivial exponent $\alpha^\prime =0.67\pm 0.01$  is obtained.

To justify the possible effect  of the hidden TCP on the CEP, as suggested
in\cite{Hatta:2003PRD,Schaefer:2006}, we analyze the behavior of the specific heat around
the TCP.  We find nontrivial critical exponents $\alpha=0.40\pm0.01$ and
$\alpha=0.45\pm0.01$,  for SU(2) and SU(3) cases, respectively.
This result, in spite of being close, is not in agreement with the respective mean field
value ($\alpha=1/2$). However, they can justify the crossing effect observed.
We notice that the closest distance between the TCP and the CEP in the phase diagram
occurs in the T-direction ($(T^{TCP} - T^{CEP})<(\mu_B^{CEP} - \mu_B^{TCP})$), and is
more clear in the SU(2) case.

\section{Summary}

We verified that our  model calculation  reproduces qualitative features of the QCD phase
diagram at $\mu_B=0$:
for $m_i=0$ the
chiral transition is second-order  for $N_f = 2$ and  first-order for $N_f\geq3$. Using
realistic values for the current quark masses we find the location of the CEP in both
SU(2) and SU(3) NJL models.

It was shown that the baryon number susceptibility and the specific heat diverge at the
CEP. The critical exponents for $\chi_B$ around the CEP, in both $N_f=2$ and $N_f=3$ NJL
models, are consistent with the mean field values $\epsilon=\epsilon'=2/3$.
For the specific heat, the nontrivial values  of $\alpha$ ($1/2<\alpha<2/3$) around the 
CEP can be interpreted as a crossover from a mean field tricritical exponent ($\alpha=1/2$) 
to an Ising-like critical exponent ($\alpha=2/3$).

A better insight to the difficult task of the analysis of the phase diagram of QCD can be
provided by an extension of the NJL model where quarks interact with the temporal gluon
field represented by the Polyakov loop dynamics.

\begin{flushleft} \textbf{Acknowledgments}
\end{flushleft}

Work supported by grant SFRH/BPD/23252/2005 from F.C.T. (P. Costa), Centro de F\'{\i}sica
Te\'orica and FCT under project POCI 2010/FP/63945/2005.



\begin{thebibliography}{}

\bibitem{Asakawa:1989NPA}
        M. Asakawa {\it et al},
        Nucl. Phys. {A {\bf 504}}, 668 (1989).

\bibitem{Hatta:2003PRD}
        Y. Hatta {\it et al},
        Phys. Rev. D {\bf 67}, 014028 (2003).

\bibitem{Schaefer:2006}
        B.-J. Schaefer {\it et al},
        Phys. Rev. D {\bf 75}, 085015 (2007).

\bibitem{Costa:2007PLB}
				 P. Costa {\it et al},
		     Phys. Lett. B {\bf 647},431 (2007); 
		     P. Costa {\it et al}, 	
		     arXiv:0801.3417v1 [hep-ph].

\bibitem{Fodor:2004JHEP}
        Z. Fodor {\it et al},
        J. High Energy Phys. 0204, 050 (2004).

\bibitem{Kunihiro:PR1994}
        T. Hatsuda {\it et al},
        Phys. Rept. {\bf 247}, 221 (1994).

\bibitem{Rehberg:1995PRC}
        P. Rehberg {\it et al},
        Phys. Rev. C {\bf 53}, 410 (1996).

\bibitem{Costa:2003PRC}
        P. Costa {\it et al},
        Phys. Rev. C {\bf 70}, 025204 (2004).

\bibitem{Costa:2005PRD}
        P. Costa {\it et al},
        Phys. Rev. D {\bf 70}, 116013 (2004);
        Phys. Rev. D {\bf 71}, 116002 (2005);
        Phys. Lett. B {\bf 560}, 171 (2003);
        Phys. Lett. B {\bf 577}, 129 (2003).

\bibitem{Pisarski:1984PRD}
        R.D. Pisarski {\it et al},
        Phys. Rev. D {\bf 29}, 338 (1984).

\end{thebibliography}
\end{document}